\documentclass{kluwer} 
\usepackage{psfig}
\begin{document}                                                                                   
\begin{article}
\begin{opening}         
\title{Gas and Dust in Protogalaxies}

\author{Francoise \surname{Combes}}  
\runningauthor{F. Combes}
\runningtitle{Gas and Dust in Protogalaxies}
\institute{DEMIRM, Observatoire de Paris, \\
61 Av. de l'Observatoire, F-75 014, Paris, France}
\date{September 1, 1999}

\begin{abstract}
The study of high-redshift objects is  rapidly developing, 
allowing to build the star formation history of the Universe.
Since most of the flux from starbursts comes out in the FIR
region, the submm and mm are privileged domains for
the exploration at high z. I review the recent work on
galaxies at high redshift in this wavelength region, for
the continuum as well as for the line detection (dust and
molecular gas). Perspectives are discussed to detect
early objects (maybe protogalaxies) with the future
large millimeter instruments.
\end{abstract}
\end{opening}           

\section{Introduction}

Considerable progress has been made in our knowledge of
galaxy evolution from the faint galaxy studies made possible
by HST deep imaging, ground-based spectroscopy, and
wide faint galaxy surveys (e.g. Ellis, \cite*{ellis98}; Steidel
et al. \cite*{steidel96}; Shade et al. \cite*{shade96}).
This progress has been led by the optical domain,
but crucial information came also from the far-infrared
and millimeter domains: the cosmic IR and submm background 
radiation discovered by COBE (e.g. Puget et al. \cite*{puget96};
Hauser et al. \cite*{hauser98}) yields an insight on the global
past star-formation of the Universe, and the sources discovered
at high redshift in the millimeter continuum and lines 
yield information on the structure of the past starbursts
(Smail et al. \cite*{smail97}; Guilloteau et al. \cite*{guillot99}).
 From all these data, a global view of star formation
as a function of lookback time has been derived
(e.g. Madau et al. \cite*{madau96}; Glazebrook et al. 
\cite*{glaze99}), which still is submitted
to big uncertainties, especially at high redshift.
In particular, it is possible that the optically derived
star formation rate is under-estimated, due to dust
obscuration, and that only infrared/submm surveys could give
the correct information (Guiderdoni et al. \cite*{guider97}).

This review focus on the dust and molecular content of galaxies,
as a way to trace the evolution of star formation, and
to detect the location of starbursts as a function
of redshift. First the present state of knowledge is
detailed, concerning CO emission lines as well as dust
continuum, and their interpretation is discussed 
(respective role of starburst and AGN for instance).
 Then perspectives are drawn concerning the future surveys
that will be conducted with the next generation of millimeter
instruments.

\section{Detection of Molecular Gas at High Redshift}  

\subsection{Emission lines}
 The detection of high-redhifted ($z>2$) millimeter CO lines in the
hyperluminous object IRAS 10214+4724
($z=2.28$, Brown \& Vanden Bout 1992, Solomon et al. 1992), has opened
a new way of research to tackle the star formation history of the Universe.
 Although the object turned out to be highly gravitationally amplified, it
revealed however that galaxies at this epoch could have large
amounts of molecular gas, excited by an important starburst,
and sufficiently metal-enriched to emit detectable CO emission lines.
 The latter bring fundamental information about the cold gas component
in high-z objects and therefore about the physical conditions of
the formation of galaxies and the first generations of stars.
At high enough redshifts, most of the galaxy mass could be molecular.
 The main problem to detect this molecular component could be its low
metallicity, but theoretical calculations have shown that in a violent
starburst, the metallicity could reach solar values very quickly
(Elbaz et al. 1992).

After the first discovery, many searches for other candidates took place,
but they were harder than expected, and only a few,
often gravitationally amplified,
objects have been detected: the lensed Cloverleaf quasar
H 1413+117 at $z=2.558$ (Barvainis et al. 1994),
the lensed radiogalaxy MG0414+0534 at $z=2.639$ (Barvainis et al. 1998),
the possibly magnified object
BR1202-0725 at $z=4.69$ (Ohta et al. 1996, Omont et al. 1996a),
the amplified submillimeter-selected hyperluminous galaxies SMM02399-0136
at $z=2.808$ (Frayer et al. 1998), and SMM 14011+0252 at 2.565
(Frayer et al. 1999), and the magnified BAL quasar APM08279+5255,
at $z=3.911$, where the gas temperature derived from the CO lines is
$\sim$ 200K, maybe excited by the quasar (Downes et al. 1999a).
Recently Scoville et al. (1997b) reported the detection of the first
non-lensed object at $z=2.394$, the weak radio galaxy 53W002,
and Guilloteau et al. (1997) the radio-quiet quasar BRI 1335-0417, at $z=4.407$,
which has no direct indication of lensing.
If the non-amplification is confirmed, these objects
 would contain the largest molecular contents known
($8-10 \cdot 10^{10}$ M$_\odot$ with a standard CO/H$_2$
conversion ratio, and even more
if the metallicity is low).
The derived molecular masses are so high that H$_2$ would constitute
between 30 to 80\% of the total dynamical mass (according to the unknown
inclination), if the standard CO/H$_2$ conversion ratio was adopted.
The application of this conversion ratio is however doubtful, and it is
possible that the involved H$_2$ masses are 3-4 times lower (Solomon
et al. 1997).

\begin{figure}
\centering
\psfig{figure=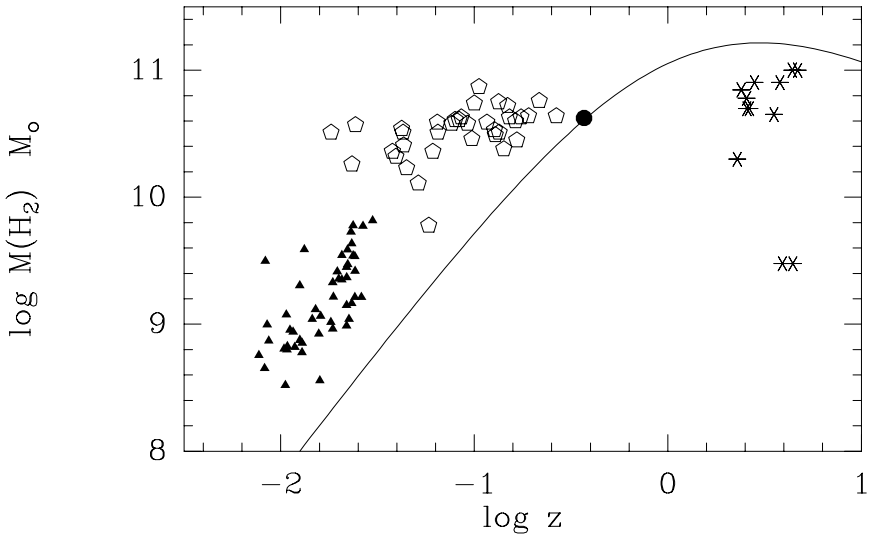,bbllx=2cm,bblly=1cm,bburx=12cm,bbury=16cm,width=12cm,angle=-90}
\caption{H$_2$ masses for the CO-detected objects at high redshift 
(stars), compared to the ultra-luminous-IR sample of Solomon et al. (1997,
open pentagons), to the Coma supercluster sample from Casoli et
al (1996, filled triangles), and to the quasar
3c48, marked as a filled dot (Scoville et al. 1993, Wink et al. 1997). The curve
indicates the 3$\sigma$ detection limit of I(CO) = 1 K km/s at the IRAM-30m
telescope (equivalent to an rms of 1mK, with an assumed $\Delta V$ = 300km/s).
Note the absence of detected objects between  0.36 and 2.2 in redshift,
where the sensitivity is insufficient, and the gravitational 
lenses maybe not yet frequent enough to compensate.  
The points at high $z$ can be detected well below the sensitivity
 limit, since they are gravitationally amplified. }
\label{emis}
\end{figure}

The CO line detections at high $z$ up to now are summarized in 
Table \ref{COdata}, and 
the molecular masses as a function of redshift are
displayed in Fig. \ref{emis}. It is clear from this figure that
our present sensitivity prevents detection of CO lines 
above a redshift of 0.4, unless the objects are lensed;
 but this will rapidly change with
the new millimeter instruments planned over the world
(the Green-Bank-100m of NRAO, the LMT-50m of UMass-INAOE,
the ALMA (Europe/USA) and the
LMSA (Japan) interferometers). It is therefore interesting to
predict with simple models the detection capabilities, as a function
of redshift, metallicity or physical conditions in the high-z objects.
In particular, it would be highly interesting to detect not only
the few exceptional amplified monsters in the sky, but also the
widely spread normal galaxy population of the young universe.
A previous study of galaxies at very high redshift (up to $z=30$) 
by Silk \& Spaans \cite{silk97} concluded
that CO lines could be even more easy to detect than the continuum;
The models presented in section 4 do not agree with this conclusion.

Today galaxies are detected in the optical up
to $z=$ 6, when the age of the universe is about
5\% of its age, or 10$^{10}$ yr in a standard flat universe model. 
For larger redshifts, it is likely that the total amount of
cumulated star formation is not a significant fraction of the total
(e.g. Madau et al. 1996). However, it is of overwhelming interest
to trace the first star-forming structures, as early  as possible
to constrain theories of galaxy formation.

\begin{table}[h]
\caption[ ]{CO data for high redshift objects}
\begin{flushleft}
\begin{tabular}{lllclcl}  \hline
Source    & $z$   &  CO  & S  & $\Delta$V& MH$_2$   & Ref  \\
                &       &line  & mJy  & km/s & 10$^{10}$ M$_\odot$    &           \\
\hline
F10214+4724  & 2.285 & 3-2  & 18 & 230  & 2$^*$      &  1   \\
53W002          & 2.394 & 3-2  &  3 & 540  & 7              &  2   \\
H 1413+117    & 2.558 & 3-2  & 23 & 330  & 2-6 $^*$  &  3   \\
SMM 14011+0252&2.565& 3-2  & 13 & 200  & 5$^*$    &  4   \\
MG 0414+0534& 2.639 & 3-2  &  4 & 580  & 5$^*$       &  5   \\
SMM 02399-0136&2.808& 3-2  &  4 & 710  & 8$^*$      &  6   \\
6C1909+722     &3.532& 4-3  &  2 & 530  & 4.5           &  7   \\
4C60.07           &3.791& 4-3  &  1.7 & 1000  & 8           &  7   \\
APM 08279+5255&3.911& 4-3  &  6 & 400  & 0.3$^*$    &  8   \\
BR 1335-0414& 4.407 & 5-4  &  7 & 420  & 10              &  9   \\
BR 0952-0115& 4.434 & 5-4  &  4 & 230  & 0.3$^*$      &  10   \\
BR 1202-0725& 4.690 & 5-4  &  8 & 320  & 10              &  11   \\
\hline 
\end{tabular}
\end{flushleft}
$^*$ corrected for magnification, when estimated\\
Masses have been rescaled to $H_0$ = 75km/s/Mpc. When multiple images
are resolved, the flux corresponds to their sum\\
(1) Solomon et al. (1992), Downes et al. (1995); (2) Scoville et al. (1997b); 
(3) Barvainis et al. (1994, 1997); (4) Frayer et al. (1999);
(5) Barvainis et al. (1998); (6) Frayer et al. (1998); 
(7) Papadopoulos et al. (1999);
(8) Downes et al. (1999a); (9) Guilloteau et al. (1997); 
(10) Guilloteau et al. (1999); (11) Omont et al. (1996a)
\label{COdata}
\end{table}

\vspace*{-0.5cm}
\subsection{Absorption lines}

\begin{figure}
\psfig{width=12cm,file=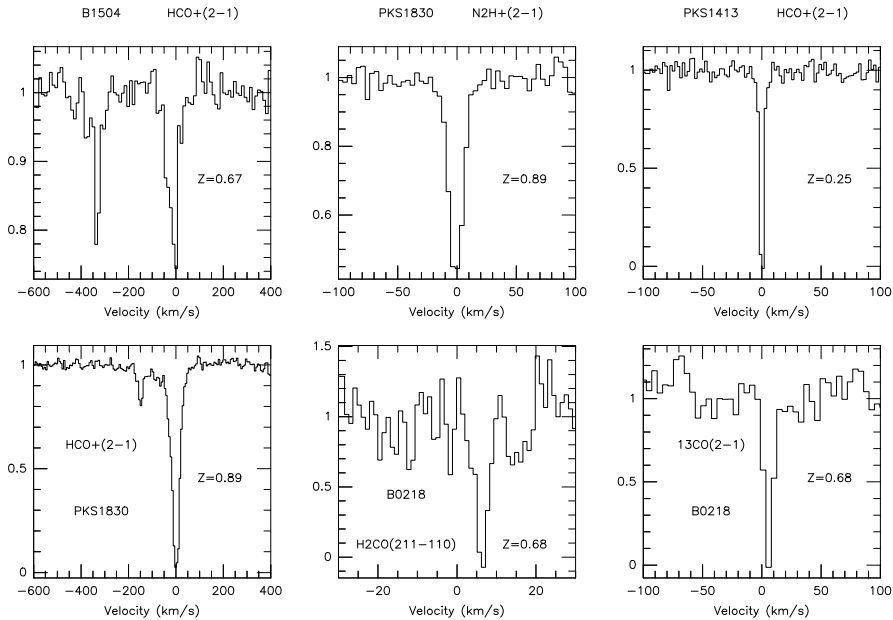,bbllx=2cm,bblly=1cm,bburx=20cm,bbury=26cm,angle=-90}
\caption{ Examples of molecular absorption lines detected in the millimeter
range. Lines can be extremely narrow (below 1km/s) up to quite broad
(100km/s). More than 20 different molecules or transitions have been detected in one
single absorption system. Here the continuum sources are 
B3 1504+377 (Wiklind \& Combes 1996b), PKS 1830-211 (Wiklind \& Combes 1996a),
PKS 1413+357 (Wiklind \& Combes 1997) and B0218+357 (Wiklind \& Combes 1995).
The signal has been normalised to the continuum level detected.}
\label{absor}
\end{figure}

Molecular absorption lines are also a powerful tool to study
the interstellar medium of galaxies at high redshift
(e.g. Combes \& Wiklind \cite*{com96}).
 A sample of the molecular lines detected is shown in Fig. \ref{absor}.

These molecular absorption objects are the continuation
at high column densities (10$^{21}$--10$^{24}$ cm$^{-2}$)
of the whole spectrum of absorption systems, from the
Ly$\alpha$ forest (10$^{12}$--10$^{19}$ cm$^{-2}$)
to the damped Ly$\alpha$ and HI 21cm absorptions
(10$^{19}$--10$^{21}$ cm$^{-2}$). It is currently thought
that the Ly$\alpha$ forest originates from gaseous filaments
in the extra--galactic medium, that the damped and HI
absorptions involve mainly the outer parts of spiral
galaxies. The molecular absorptions concern the central
parts of galaxies.
The properties of molecular absorptions detected in the millimeter
domain so far, are summarised in Table \ref{tababs}.

\begin{table}[h]
\caption{Properties of molecular absorption line systems in the mm}
\begin{center} 
\begin{tabular}{lcccccrc}
Source & z$_{\rm a}^{a}$ & z$_{\rm e}^{b}$ &
$N_{\rm CO}$ & $N_{\rm H_2}$ & $N_{\rm HI}$$^{e}$ &
A$_{V}^{\prime c}$ & $N_{\rm HI}/N_{H_2}$ \\
 & & & cm$^{-2}$ & cm$^{-2}$ & cm$^{-2}$ & & \\
\hline \\
Cen-A   & 0.00184 & 0.0018  & $1.0 \cdot 10^{16}$ & $2.0 \cdot 10^{20}$
& $1.0 \cdot 10^{20}$ & 50 & 0.5 \\
3C 293   & 0.0446 & 0.0446 & $\geq 3.0 \cdot 10^{16}$ & $\geq 6.0 \cdot 10^{20}$
& $1.2 \cdot 10^{21}$ & -- & $\geq$ 0.5 \\
PKS1413+357   & 0.24671 & 0.247  & $2.3 \cdot 10^{16}$ & $4.6 \cdot 10^{20}$
& $1.3 \cdot 10^{21}$ & 2.0 & 2.8 \\
B3\,1504+377A & 0.67335 & 0.673  & $6.0 \cdot 10^{16}$ & $1.2 \cdot 10^{21}$
& $2.4 \cdot 10^{21}$ & 5.0 & 2.0 \\
B3\,1504+377B & 0.67150 & 0.673   & $2.6 \cdot 10^{16}$ & $5.2 \cdot 10^{20}$
& $<7 \cdot 10^{20}$ & $<$2 & $<$1.4 \\
B\,0218+357   & 0.68466 & 0.94   & $2.0 \cdot 10^{19}$ & $4.0 \cdot 10^{23}$
& $4.0 \cdot 10^{20}$ & 850 & $1 \cdot 10^{-3}$ \\
PKS1830--211A & 0.88582 & 2.51      & $2.0 \cdot 10^{18}$ & $4.0 \cdot 10^{22}$
& $5.0 \cdot 10^{20}$ & 100 & $1 \cdot 10^{-2}$ \\
PKS1830--211B & 0.88489 & 2.51      & $1.0 \cdot 10^{16 d}$ &
$2.0 \cdot 10^{20}$ & $1.0 \cdot 10^{21}$ & 1.8 & 5.0 \\
PKS1830--211C & 0.19267 & 2.51      & $<6 \cdot 10^{15}$                   &
$<1 \cdot 10^{20}$ & $2.5 \cdot 10^{20}$ & $<$0.2 & $>$2.5 \\
\hline
\end{tabular}
\end{center}
$^{a}${Redshift of absorption line}
$^{b}${Redshift of background source}
$^{c}${Extinction corrected for redshift using a Galactic extinction law}
$^{d}${Estimated from the HCO$^{+}$ column density
of $1.3 \cdot 10^{13}$\,cm$^{-2}$}
$^{e}${21cm HI data taken from Haschick \& Baan (1985, 3C293) and Carilli et al. 1992,
1993, 1998}
\label{tababs}
\end{table}

The utility of molecular absorption lines comes from the
high sensitivity. Due to the small extent of the background continuum
source, the signal is not diluted, there is no distance dependence.
Molecular absorption lines are as easy to detect
at z$\approx$1 as at z$\approx$5, provided the 
continuum sources exist. Then, direct opacity are measured,
and it is almost as easy to detect many high dipole molecules,
(HCO$^+$ or HCN) as CO. 

About 15 different molecules have been detected
in absorption at high redshifts, in a total of 30
different transitions. This allows a detailed
chemical study and comparison with local clouds.
Within the large dispersion in column densities,
and in molecular cloud properties, the
high redshift systems do not appear to be different from local
ones, suggesting that the conditions for star formation are
the same up to z$\sim$1 as at the present.

 A systematic survey for absorption lines has also been 
undertaken in front of about a hundred continuum sources 
candidates, selected from flat--spectrum continuum
sources. The continuum needs to be at least 0.2 Jy to allow
detection of intervening molecular gas.
The redshift of the absorbing candidate is known, either
from previously detected HI absorption,
or from optical lines emission. When the continuum source is strong
enough, at least 1 Jy, and no redshift is known, it is possible
to search for absorption lines by scanning in frequency 
(cf. Wiklind \& Combes \cite*{wik96a}).
This last method is the most promising with the new
generation millimeter instruments, that will gain
an order of magnitude in sensitivity. Indeed, the best
candidates are the most obscured ones, where no redshift
is available.

A lot more absorption systems could be found with the future
instruments, at faint continuum flux, since the number counts of quasars
are a non-linear function of flux and the local
luminosity function is steep (see Peacock \cite*{peacock85}).
However, the flat-spectrum radio-loud quasars distribution
is decreasing sharply at $z > 3$ (Shaver et al. \cite*{shaver96}),
and this is not likely due to obscuration, since they are 
radio-selected quasars. If quasars are associated to galaxy
formation and interactions, this tends to show that the decrease of 
star-formation rate beyond $z = 3$ is real and not an extinction effect.
The relative absence of strong continuum radio sources at high 
redshift will not help the tracing of protogalaxies 
through absorption techniques.

\section{Detection of Dust at High Redshift}

\subsection{Submm continuum surveys}

\begin{figure}
\psfig{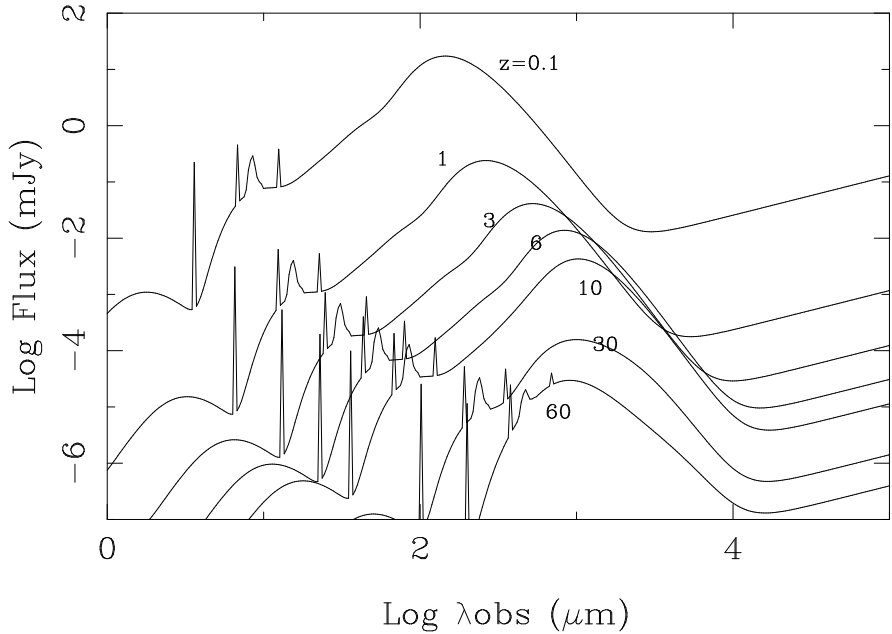}
\caption{ Spectral flux distribution for a typical ULIG starburst source in the radio
and far-infrared, for various  redshifts $z$ = 0.1, 
1, 3, 6, 10, 30, 60 ($H_0$ = 75km/s/Mpc, $q_0$ = 0.5).
 At right is a synchrotron spectrum, in a power-law of slope -0.7,
and left the emission from dust, modelled by PAHs, very small
grains and big grains, as in D\'esert et al. (1990) to fit the Milky Way
data. It has been assumed here that the dust properties
are the same as in our Galaxy, and that the power of the starburst 
is the same at any redshift, i.e. $T_{\rm dust}^6 - T_{\rm bg}^6$ is conserved.}
\label{sed}
\end{figure}

The spectral energy distribution (SED) of galaxies over the radio, 
mm and FIR domains has a characteristic maximum around 60-100 $\mu$m
due to dust heated by newly born stars, and the interstellar radiation
field (see Fig. \ref{sed}).
This maximum depends on the dust temperature, and the curve is that 
of a grey-body, where the optical thickness can be modelled by a power-law
in frequency, $\tau \propto \nu^\beta$, where $\beta \sim 1.5-2$,
according to the nature of dust. In the Rayleigh-Jeans domain 
(frequencies lower than the maximum), the flux increases almost as
$\nu ^4$, and this creates what is called a negative K-correction, 
i.e. it begins to be more easy to detect objects at high redshift than
low redshift, at a given frequency, and sky surveys could be dominated 
by remote objects (see e.g. Blain \& Longair \cite*{blain93, blain96}). The source counts
could be inverted in favor of high-redshift objects, if they exist in equal 
numbers (not depopulated by strong evolution effects). The millimeter domain
becomes then a privileged tool to tackle galaxy formation.

In Fig. \ref{sed}, we see that at low frequency (lower than 100 GHz, or
3mm in wavelength), the radio spectrum is due to synchrotron processes. 
There is a marked flux minimum that can be used as a redshift indicator
(e.g. Carilli \& Yun \cite*{car99}; Blain \cite*{blain99a}). 
Indeed, there is a well-known tight correlation
between the synchrotron radio power and the far-infrared emission in 
star-forming galaxies (Condon \cite*{condon92}). This correlation is thought to
arise because both FIR and non-thermal radio emission are both 
directly proportionnal to recent star-formation (the FIR being a good
measure of massive stars luminosity, and the radio of the rate of supernovae).
However, the z-indicator is somewhat ambiguous, since there is a
degeneracy between increasing the redshift or decreasing the dust 
temperature. 

The bulk of the high redshift galaxies presently known have
been discovered in optical. But these surveys could have
missed dust-enshrouded starbursts, since we now know
that dust and high metallicity occur very early in the universe
(cf. the previous sections). Submm and FIR deep surveys are then the best
strategy to detect starbursting proto-galaxies, and many
such works have been undertaken, either with sensitive 
array-bolometers on single dishes (IRAM-30m, SCUBA on JCMT, ...)
mm-interferometers, or with ISOPHOT and ISOCAM 
on board of ISO satellite.  

The first deep search was made with the SCUBA bolometer
(Holland et al. \cite*{holland99}) towards  a cluster of galaxies, thought to serve as a
gravitational lens for high-$z$ galaxies behind (Smail et al. 1997). The
amplification is in average a factor 2. This has the combined advantage
to increase the sensitivity, and to reduce the source confusion, since
there should be little contamination from cluster galaxies (Blain \cite*{blain97}).
A large number of sources were found,
all at large redshifts ($z > 1$), extrapolated
to 2000 sources per square degree (above 4mJy),
revealing a large positive evolution with redshift, i.e. an increase of
starbursting galaxies. Searches toward the Hubble Deep Field-North
(Hughes et al. 1998), and towards the Lockman hole and SSA13 (Barger et al. \cite*{barger98}),
have also found a few sources, allowing to derive a similar density of sources:
800 per square degree, above 3 mJy at 850 $\mu$m. This already can account for
50\% of the cosmic infra-red background (CIRB), that has been estimated by
Puget et al. (1996) and Hauser et al. (1998) from COBE data.
The photometric redshifts of these sources range between 1 and 3.
Their identification with optical objects might be  uncertain (Richards \cite*{rich99}).
However, Hughes et al. (1998) claim that the star formation rate derived from
the far-infrared might be in some cases 10 times higher than derived
from the optical, due to the high extinction. 

Eales et al. (1999) surveyed some of the CFRS fields at 850$\mu$m
 with SCUBA and found also that the sources can account for a significant
fraction of the CIRB background ($\sim$ 30\%). Their interpretation in terms
of the star formation history is however slightly different, in that they
do not exclude that the submm luminosity density could evolve in the same
way as the UV one.
Deep galaxy surveys at 7 and 15$\mu$m with ISOCAM also see an evolution with
redshift of star-forming galaxies: heavily extincted starbursts represent less
than 1\% of all galaxies, but 18\% of the star formation rate out to $z = 1$
(Flores et al. 1999).

Now that a few dozens of submm sources have been catalogued
(Barger et al. \cite*{barger99a}, Smail et al. 1999), the count rates are confirmed,
i.e. $\sim$ 1000 source per square degree, above 3 mJy, at 850 $\mu$m,
and even 8000 above 1 mJy, from the gravitationally amplified
cluster fields (Blain et al. \cite*{blain99b}). The cumulative count
rate can be fitted by a power-law, above 2 mJy, with a slope
of -2.2. The main difficulty appears to be the identification
of the submm sources with optical or radio counterparts:
the spatial resolution of the submm surveys are several arcsecs, 
with sometimes systematic uncertainties, and some of
the previous claimed identifications have been reconsidered
(e.g. Barger et al. \cite*{barger99b}, Downes et al. \cite*{downes99b}).
Follow-up in the radio (CO lines) or near-infrared, or
optical to find redshifts, are much slower than the 
surveys themselves. At least 20\% of the sources reveal an
AGN activity, and the bulk of the sources are at relatively
low redshift $ 1 < z < 3 $ (Barger et al. \cite*{barger99a}).

\subsection{Extremely red objects, EROs}

To search for primeval galaxies, already Elston et al. \cite{elston88}
had identified extremely red objects that are conspicuous only in the
near-infrared, and have R - K colors larger than 5. The ERO class
could include several categories of objects: essentially 
old stellar populations at high redshift, or high-$z$ dust-reddened
starbursts (Hu \& Ridgway \cite*{hu94};  Cowie et al. \cite*{cowie96}).
Maybe 10\% of the submm sources could be EROs
(Smail et al. \cite*{smail99}). A proto-typical ERO at $z$=1.44 
(Dey et al. \cite*{dey99}) has been detected in submm continuum 
(Cimatti et al. \cite*{cimatti98}), and has been found to be
an ultraluminous starburst shrouded by dust,
with a star formation rate of 200-500 M$_\odot$ per yr.
 The surface density of such EROs at K $<$ 20 and
color R - K $>$ 6 is about 500 per square degree,
and for R - K $>$ 7, about 50 per square degree
(which is comparable to the number of QSO with
B $<$ 21.5, Andreani et al. \cite*{andreani99}).
 At least for the few objects known, the AGN cannot
be the source of the huge luminosity, given
the FIR/radio ratio.

\subsection{AGN or Starbursts}

For a long time, the highest redshift objects known were
quasars, although now, with higher sensitivities, the
situation is reversed (e.g. Hu et al. \cite*{hu98}).
Surveys have therefore been done in the millimeter 
continuum and lines for high-z quasars (McMahon et al. \cite*{mcmahon94};
Isaak et al. \cite*{isaak94}; Omont et al. \cite*{omont96a, omont96b};
Guilloteau et al. \cite*{guillot99}). Results have shown that
the far-infrared luminosities of many of them are due
to dust heated by a starburst, although the AGN activity
is simultaneously present. But since the manifestations
of the two nuclear activities (starburst or AGN) are in many
cases similar, and are most of the time associated,
it has become a controversial question to 
disentangle the two interpretations. Ultra-luminous
starbursts take place over the few central 100pc (Solomon
et al. \cite*{sol97}), and even pure AGN activities can be
mimicked by radio supernovae (Boyle \& Terlevich \cite*{boyle98}).
Genzel et al. \cite*{genzel98} from ISO mid-infrared diagrams
concluded that the ultra-luminous galaxies are powered at
30\% from AGN and 70\% from star formation; a similar
conclusion is suggested by Cooray \& Haiman \cite{cooray99}
for submm catalogued sources. McMahon et al. \cite{mcmahon99}
however suggest that a much more significant fraction of
submm sources, between 15 to 100\%, could be AGN-powered.
Note that distinction criteria
are hard to find (cf. Stein \cite*{stein95}), and even symbiotic
starburst-black hole models are likely (Williams et al. \cite*{will99}).
Discriminating between the two possibilities
can have important consequences on the star 
formation history of the Universe.   

\subsection{ Star Formation Rate}

One of the breakthrough due to recent progress on faint 
galaxies has been the inventory of the amount of star
formation at every epoch (e.g. Madau et al. \cite*{madau96}).
 The comoving star formation rate is increasing like
$(1+z)^4$ from $z = 0$ to $z = 1$, and then decreases again
to the same present value down to $z = 5$. But this relies
on the optical studies, i.e. on the UV-determined star
forming rates in the rest-frame. If early starbursts are dusty,
this decrease could be changed into a plateau
(Guiderdoni et al. \cite*{guider97, guider98}; Blain et al. \cite*{blain99c}).
Since the AGN-starbursts nature of the submm 
sources is still an open question,
Trentham et al. \cite{trentham99} consider two extreme 
possibilities: either the submm faint galaxies are all
dusty starbursts, but even then, they do not dominate
the star formation rate in the Madau plot at any redshift;
or they are dusty AGN, and again they cannot represent
more than a few percents of the present density of 
dark objects, as inferred by Magorrian et al. \cite{mago98}.

\section{Perspectives with Future mm Instruments}

 If progenitors of quasars or protogalaxies form at
high redshifts (larger than 10), then the millimeter domain 
is the best place to detect them (Loeb \cite*{loeb93};
Braine \cite*{braine95}). Both continuum and
line emission could be detected, provided enough sensitivity,
with about 10 times more collecting surface as present 
ones (either with a single dish, as the GBT-100m of NRAO,
or with the ALMA interferometer).
As was clear in the previous sections, the detection of
the submm continuum emission from actively star-forming 
objects at high redshifts is much easier than the CO line detection.
The line emission does not have such a negative
K-correction, since in the low frequency domain, the
flux of the successive lines increases roughly as $\nu^2$
(optically thick domain), instead of  $\nu^4$ for the 
continuum. 
Nevertheless the line emission is essential to study
the nature of the object (the AGN-starburst connection
for instance), and deduce more physics (kinematics,
abundances, excitation, etc..). Given the gas and dust 
temperatures, the maximum flux is always reached at
much lower frequencies than in the continuum, since the lines
always reflect the energy difference between two levels; 
this is an advantage, given the largest atmospheric 
opacity at high frequencies.

\subsection{Predicted Line and Continuum Fluxes}

The best tracer of molecular gas at large-scale is the CO
molecule, the most abundant after H$_2$. All other
molecules will give weaker signals. The fine-structure CII line at
158 $\mu$m, formed in PDR at the border of molecular clouds,
is also thought to be a useful tool for proto-galaxies
or proto-quasars (Loeb \cite*{loeb93}), but it has
revealed disapointing in starbursts, or compact
objects, due to optical thickness or inefficiency of 
gas heating (e.g. Malhotra et al.
\cite*{mal97}): the L$_{\rm CII}$/L$_{\rm FIR}$ ratio decreases
as L$_{\rm FIR}$/L$_{\rm B}$ increases.

To model high-redshift starburst objects, let us
extrapolate the properties of more local ones:
the active region is generally confined to a compact nuclear
disk, sub-kpc in size (Scoville et al. \cite*{sco97a},
Solomon et al. \cite*{sol90, sol97}). The gas is much denser
here than in average over a normal galaxy, of the order
of 10$^4$ cm$^{-3}$, with clumps at least of  10$^6$ cm$^{-3}$
to explain the data on high density tracers (HCN, CS..);
large gas masses can pile up in the center, due to
torques exerted in galaxy interactions and mergers
(e.g. Barnes \& Hernquist \cite*{barnes92, barnes96}).
To schematize, the ISM maybe modelled by two
density and temperature components, at 30 and 90K  
(cf. Combes et al. \cite*{com99}). The total molecular mass
considered will be $6 \cdot 10^{10}$ M$_\odot$ and the 
average column density N(H$_2$) of 10$^{24}$ cm$^{-2}$,
typical of the Orion cloud center. 

Going towards high redshift ($z > 9$), the temperature of
the cosmic background T$_{\rm bg}$ becomes of the same order
as the interstellar dust temperature, and the excitation of
the gas by the background radiation competes with that
of gas collisions. It might then appear easier to detect the lines
(Silk \& Spaans \cite*{silk97}), but this is not the case
when every effect is taken into account. To have an idea
of the increase of the dust temperature with $z$, the simplest
assumption is to consider the same heating power due
to the starburst. At a stationary state, the dust must then radiate
the same energy in the far-infrared that it receives from
the stars, and this is proportional to the quantity
  $T_{\rm dust}^6 - T_{\rm bg}^6$, if the dust is optically thin, and
its opacity varies in $\nu^{\beta}$, with $\beta$ = 2. Keeping this 
quantity constant means that the energy re-radiated by the dust,
proportional to  $T_{\rm dust}^6$, is always equal to the energy it
received from the cosmic background,
proportional to  $T_{\rm bg}^6$, plus the constant energy flux
coming from the stars. Since $\beta$ can also be equal to 
1 or 1.5, or the dust be optically thick, we have also 
considered the possibility of keeping  $T_{\rm dust}^4 - T_{\rm bg}^4$
constant; this does not change fundamentally
the results.

\begin{figure}
\psfig{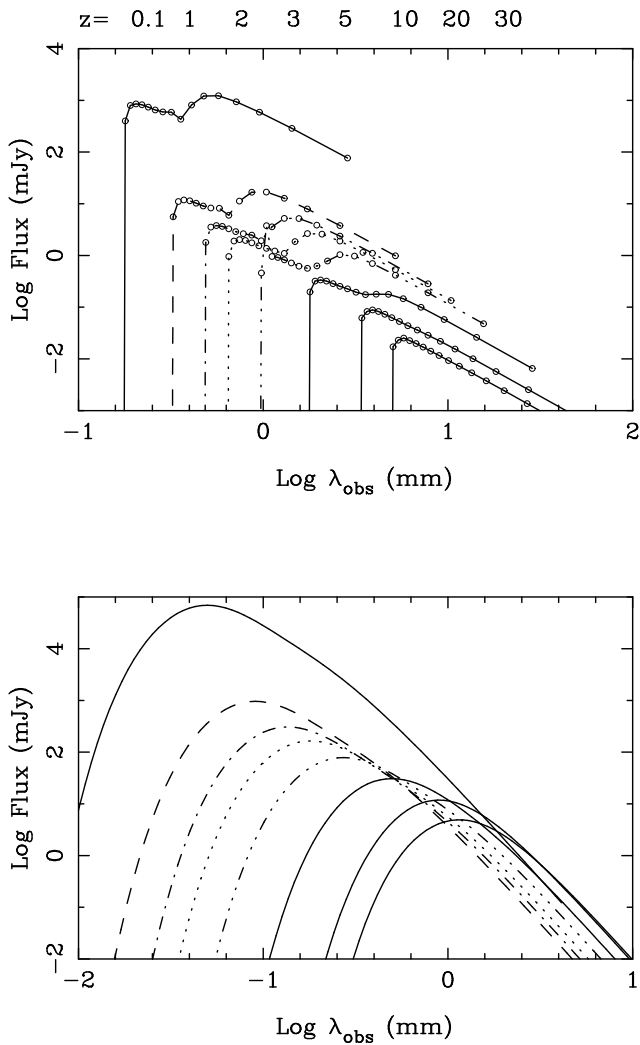}
\caption{ Expected flux for the two-component cloud model, for various
redshifts $z$ = 0.1, 1, 2, 3, 5, 10, 20, 30, and $q_0$ = 0.5.
Top are the CO lines, materialised each by a circle
(they are joined by a line only to guide the eye). 
Bottom is the continuum emission from dust. It has been assumed here 
that $T_{\rm dust}^6 - T_{\rm bg}^6$ is conserved
(from Combes et al. 1999).}
\label{cohz_39}
\end{figure}

Computing the populations of the CO rotational levels
with an LVG code, and in the case of the two component
models described earlier, the predictions of the line
and continuum intensities as a function of redshift and
frequencies are plotted in Fig. \ref{cohz_39}.

When comparing these predictions with the present 
instrumental sensitivities, it appears that the continuum is
detectable at any redshift already, for an ultra-luminous source,
while the line emission has to await the order of magnitude
increase that will be provided by the next generation
in the mm and sub-mm domains. The recent reported
detections (cf. Table \ref{COdata}) have been possible
because of gravitational lens magnifications
(or maybe for 1-2 cases, an exceptional object).

\subsection{Source Counts}

To predict the number of sources that will become available
with the future sensitivity, let us adopt a simple model of
starburst formation, in the frame of the hierarchical theory
of galaxy formation. The cosmology adopted here is
an Einstein-de Sitter model, $\Omega$ = 1, with no
cosmological constant, and $H_0$ = 75 km/s/Mpc, 
$q_0$ = 0.5.The number of mergers as a function 
of redshifts can be easily computed through the Press-Schechter
formalism (Press \& Schechter \cite*{ps74}), assuming 
self-similarity for the probability of dark halos merging 
(i.e. independent of mass, Blain \& Longair \cite*{blain93}):
the mass spectrum of bound objects is, at any redshift $z$:
$$
\frac{dN}{dM_{PS}} \propto M^{-2} \left({\frac{M}{M^*}}\right)^{\gamma/2}
\exp \left[ -\left({\frac{M}{M^*}}\right)^{\gamma}\right]
$$
where $\gamma = (3+n)/3$, and $n$ the power-law slope of the 
primordial fluctuations ($n$ = -1 here, so that 
$\gamma = 2/3$). The turn-off mass $M^*$ is increasing 
with redshift as:
$$
M^* \propto (1+z)^{-2/\gamma}
$$
Following Blain et al. \cite{blain99d}, we also write the merger
rate as a function of $M$ and $z$ under the form:
$$
\frac{d}{dt} \frac{dN}{dM_{PS}}+\phi \frac{dN}{dM_{PS}} \left(\frac{-2}{\gamma} 
\frac{dz}{dt (1+z)}\right) 
\exp \left[ (1-\alpha) \left({\frac{M}{M^*}}\right)^{\gamma}\right]
$$
This shape (with $\phi$ = 1.7  and $\alpha$ = 1.4) has been
chosen so that the merger rate is the sum of the net evolution 
of the Press-Schechter distribution at any mass, plus a specific
term, indicating that the merger probability is
maximum at equal masses, and is exponentially vanishing at 
high mass ( cf.  Blain \& Longair \cite*{blain93}). 

\begin{figure}
\psfig{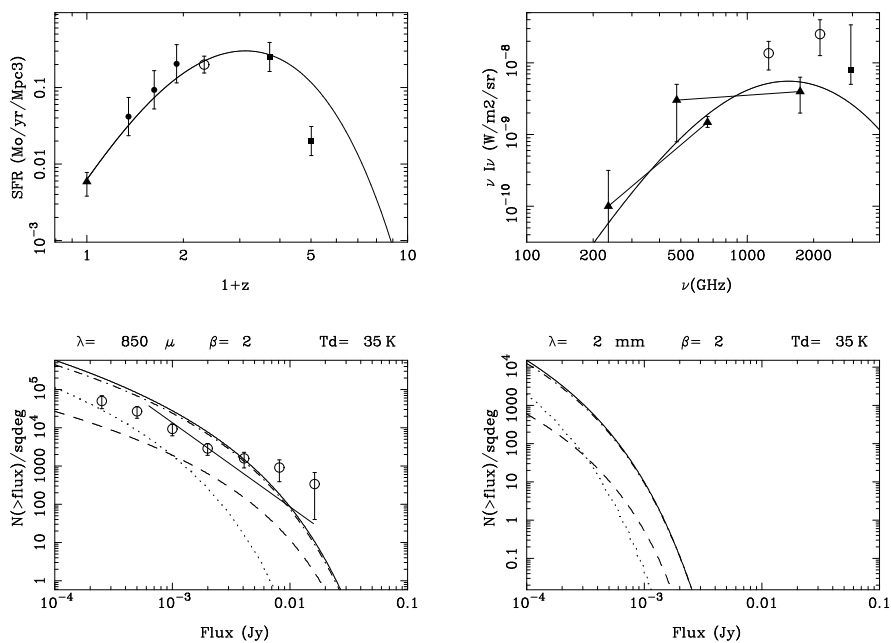}
\caption{ {\it Top left:} History of the star formation rate, adopted by 
the hierarchical model used here, compared with observations.
The point at $z=0$ is from Gallego et al. (1995), the 3 following
ones (full circles) from Flores et al. (1999), the empty
circle from Yan et al. (1999), and the two high redshift points
(filled squares), from Pettini et al. (1998).
   {\it Top right:} Predictions of the model for the far-infrared cosmic
background: the observations have been used to constrain the parameters
of the model. The filled triangles with a line joining them
symbolise the spectrum derived by Puget et al. (1996) from COBE,
the filled circles are from Hauser et al. (1998), and the 
filled square from Dwek et al. (1998). 
 {\it Bottom left and right:}  Source counts predicted at 850$\mu$m  
and 2mm respectively: the solid line is the total. Dash is the lowest
redshifts ($z < 2$); Dot-dash, intermediate ($2 < z < 5$);
Dots are the highest redshifts  ($z > 5$).
The emission from the dusty starbursts have been assumed a
grey-body at a temperature of 35K, and an opacity varying
as $\nu^2$. The empty circles are data from Blain et al. (2000),
and the straight line is a fit to the data derived by Barger et al. (1999a)
at 850$\mu$m. }
\label{scount}
\end{figure}

This gives the number of mergers at each epoch, but the 
efficiency of mergers in terms of star-formation must 
also vary considerably with redshift, with a peak at $z \sim 2$,
to agree with the observations (such as in Fig. \ref{scount}).
Also, the integration over all redshifts of the flux of all sources
should agree with the cosmic infrared background detected
by COBE (Fig. \ref{scount}). These contribute to reduce the 
number of free parameters of the modelling. To fit the source
counts, however, another parameter must be introduced 
which measures the rate of energy released in a merger
(or the life-time of the event): this rate must increase
strongly with redshift (cf. Blain et al. \cite*{blain99d}).
Once the counts are made compatible with the submm
observations, the model indicates what must be
the contributions of the various redshift classes to the
present counts (cf. Fig. \ref{scount}). It is interesting to
note that the intermediate redshifts dominate the
counts ($2 < z < 5$), if we allow the star formation
to begin before $z = 6$. At higher dust temperature, 
the counts are dominated by the highest redshifts 
($ z > 5$). But these contributions depend strongly on the adopted 
shape of the star-forming efficiency versus $z$. Observations of
these counts and their redshift distribution
will therefore bring a lot of insight in the physiscs of early
protogalaxies.

\begin{figure}
\psfig{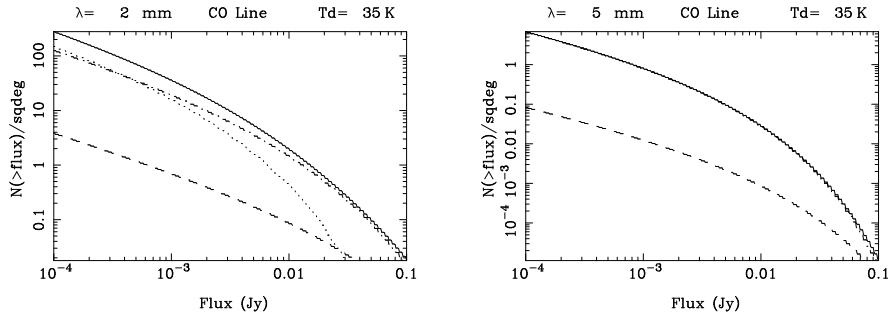}
\caption{ {\it Left} Source counts for the CO lines at an observed frequency 
of 2mm, assuming optically thick gas at T$_{ex}$ = 35K.
   {\it Right} Same for $\lambda$ = 5mm. The meanings of the line styles
are as in fig 5}
\label{scoline}
\end{figure}

It is possible to derive also the source counts for the CO line emission.
The spectral energy distribution is now obtained
with a comb-like function, representing the rotational ladder,
convolved with a Planck distribution of
temperature equal to the dust temperature, assuming the lines optically
thick. The frequency filling factor is then proportional to the 
rotational number, and therefore to the redshift, for a given observed 
frequency. The width of the lines have been assumed to be 300km/s. 
The derived source numbers are shown in Fig. \ref{scoline}.
 Note how they are also dominated
by the high redshift sources. It is however not useful to 
observe at $\lambda$ below 1mm for high-$z$ protogalaxies, but
instead to shift towards $\lambda$ = 1cm.

\section{Conclusion}

If there were as many ultra-luminous starbursts at $z=1$ than at
$z=5$, the present sub-mm continuum surveys should have detected 
them (there might still be identification problems, however).
 A good way to identify them is to search for molecular lines,
but the present instruments are not yet sensitive enough,
when they are not helped by gravitational lenses. 
 With more than an order of magnitude sensitivity
that will be available in the next decade, it will become
possible to detect normal star-forming galaxies at
high redshift, and even protogalaxies at $z > 5$.


\end{article}

\begin{thebibliography}{}

\bibitem[1999]{andreani99}Andreani P., Cimatti A., R\"ottgering H., Tilanus R.:
in  Proceedings of the Workshop 'Ultraluminous Infrared Galaxies: Monsters
    or Babies', Ringberg Castle, Germany, Sep. 1998 (astro-ph/9903121)

\bibitem[1998]{barger98}Barger, A.J., Cowie, L.L., Sanders, et al.:
1998, Nature, 394, 248

\bibitem[1999a]{barger99a}Barger, A.J., Cowie, L.L., Sanders, D.B.: 1999a, ApJ  518, L5

\bibitem[1999b]{barger99b}Barger, A.J., Cowie, L.L., Smail I. et al. : 1999b, 
AJ 117, 2656 

\bibitem[1992]{barnes92}Barnes J.E., Hernquist L.: 1992, ARA\&A 30, 705

\bibitem[1996]{barnes96}Barnes, J.E., \& Hernquist, L., 1996, ApJ, 471, 115

\bibitem[1998]{barv98}Barvainis R., Alloin D., Guilloteau S., Antonucci R.
1998, ApJ 492, L13

\bibitem[1994]{barv94}Barvainis R., Tacconi L., Antonucci R., Coleman P.: 1994, Nature
371, 586

\bibitem[1997]{blain97}Blain A.W.: 1997, MNRAS 290, 553

\bibitem[1999]{blain99a}Blain A.W.: 1999, MNRAS in press (astro-ph/9906438) 

\bibitem[2000]{blain00}Blain A.W., Ivison R.J., Kneib J-P.,  Smail I.: 2000, in Proceedings
of ``The Hy-redshift Universe'', eds. A.J. Bunker, W.J.M. van Breugel, 
(astro-ph/9908024)

\bibitem[1999c]{blain99d}Blain A.W., Jameson A., Smail I. et al. : 1999c, MNRAS in press
(astro-ph/9906311)

\bibitem[1999a]{blain99b}Blain A.W., Kneib J-P., Ivison R.J., Smail I.: 1999a, ApJ 512, L87

\bibitem[1993]{blain93}Blain A.W., Longair M.S.: 1993, MNRAS 264, 509

\bibitem[1996]{blain96}Blain A.W., Longair M.S.: 1996, MNRAS 279, 847

\bibitem[1999b]{blain99c}Blain A.W., Smail I.,  Ivison R.J., Kneib J-P.: 1999b, MNRAS 302, 632

\bibitem[1998]{boyle98}Boyle, B. J., Terlevich, R. J.: 1998, MNRAS 293, L49

\bibitem[1995]{braine95}Braine J.: 1995, A\&A 300, 20

\bibitem[1992]{brown92}Brown R., Vanden Bout P.: 1992, ApJ 397, L19

\bibitem[1998]{car98}Carilli, C.L., Menten K.M., Reid M.J., Rupen, M.P., Yun M.S.: 1998, 
ApJ 494, 175

\bibitem[1992]{car92}Carilli C.L., Perlman E.S., Stocke J.T.: 1992, ApJ 400, L13

\bibitem[1993]{car93}Carilli, C.L., Rupen, M.P., Yanny, B. 1993, ApJ 412, L59

\bibitem[1999]{car99}Carilli, C.L., Yun M.S.: 1999, ApJ 513, L13

\bibitem[1996]{cas96}Casoli F., Dickey J., Kazes I. et al.: 1996, A\&AS 116, 193

\bibitem[1998]{cimatti98}Cimatti A., Andreani P., R\"ottgering H., Tilanus R.: 
1998, Nature 392, 895

\bibitem[1999]{com99}Combes F., Maoli R., Omont A.: 1999, A\&A 345, 369

\bibitem[1996]{com96}Combes F., Wiklind T.: 1996, in "Cold gas at high redshift", ed. Breme
r M., Rottgering H., van der Werf P., Carilli C.L. (Dordrecht:Kluwer), p. 215

\bibitem[1992]{condon92}Condon, J.J.: 1992, ARAA  30, 575

\bibitem[1999]{cooray99}Cooray A.R., Haiman Z.: 1999, BAAS, 194.2007

\bibitem[1996]{cowie96}Cowie, L. L., Songaila, A., Hu, E. M., Cohen, J.G.: 1996, AJ 112, 839

\bibitem[1990]{desert90}D\'esert F-X., Boulanger F., Puget J-L.: 1990, A\&A 237, 215

\bibitem[1999]{dey99}Dey, A., Graham, J. R., Ivison, R. J. et al.: 1999, ApJ 519, 610

\bibitem[1999b]{downes99b}Downes D., Neri R., Greve A., et al.: 1999b,
A\&A 347, 809

\bibitem[1999a]{downes99a}Downes D., Neri R., Wiklind T., Wilner D.J., Shaver P.: 1999a,
ApJ 513, L1

\bibitem[1995]{downes95}Downes D., Solomon P.M., Radford S.J.E. 1995, ApJ 453, L65

\bibitem[1998]{dwek98} Dwek E. et al. : 1998, ApJ 508, 106

\bibitem[1999]{eales99}Eales, S., Lilly, S., Gear, et al.: 1999, ApJ 515, 518

\bibitem[1992]{elbaz92}Elbaz D., Arnaud M., Casse M., et al.: 1992, A\&A 265, L29

\bibitem[1998]{ellis98}Ellis R., 1998, Nature, 395, A3

\bibitem[1988]{elston88}Elston R., Rieke G.H., Rieke M.J.: 1988, ApJ 331, L77

\bibitem[1999]{evans99}Evans, A. S., Sanders, D. B., Surace, J. A., Mazzarella, J. M.: 1999,
ApJ 511, 730  

\bibitem[1999]{flores99} Flores H. et al. : 1999, ApJ 517, 148

\bibitem[1998]{frayer98}Frayer D.T., Ivison R.J., Scoville N.Z., et al., 1998, ApJ 506, L7

\bibitem[1999]{frayer99}Frayer D.T., Ivison R.J., Scoville N.Z., et al., 1999, ApJ
514, L13

\bibitem[1995]{gallego95} Gallego J., Zamorano J., Arag\'on-Salamanca A., Rego M.: 
1995, ApJ 455, L1 (erratum ApJ 459, L43) 

\bibitem[1998]{genzel98}Genzel, R., Lutz, D., Sturm, E. et al., 1998, ApJ, 498, 579

\bibitem[1999]{glaze99}
Glazebrook K., Blake C., Economou F., Lilly S., Colless M.: 1999, MNRAS 306, 843

\bibitem[1997]{guider97} Guiderdoni B., et al.: 1997, Nature 390, 257
1997, A\&A, 328, L1

\bibitem[1998]{guider98}Guiderdoni B., Hivon E., Bouchet F.R., Maffei B.: 1998 MNRAS 295, 877

\bibitem[1999]{guillot99}Guilloteau S., Omont A., Cox P., McMahon R.G., PetitJean P.: 1999,
A\&A preprint  

\bibitem[1997]{guillot97}Guilloteau S., Omont A., McMahon R.G., Cox P., PetitJean P.: 1997,
A\&A 328, L1

\bibitem[1985]{has85}Haschick, A. D., Baan, W. A.: 1985, ApJ 289, 574

\bibitem[1998]{hauser98}Hauser, M. et al., 1998, ApJ 508, 25

\bibitem[1999]{holland99}Holland W.S., Robson E.I., Gear W.K, et al.: 1999, MNRAS 303, 659

\bibitem[1998]{hu98}Hu, E.M., Cowie L.L., McMahon, R.G.: 1998, ApJ, 502 L99

\bibitem[1994]{hu94}Hu, E. M., Ridgway, S. E.: 1994, AJ 107, 1303

\bibitem[1998]{hug98}Hughes D.H., Serjeant S., Dunlop J. et al.: 1998, Nature 394, 241

\bibitem[1994]{isaak94}
Isaak, K.G., McMahon, R.G., Hills, R.E., Withington, S., 1994, MNRAS, 269, L28

\bibitem[1993]{loeb93}Loeb A.: 1993, ApJ 404, L37

\bibitem[1996]{madau96}
Madau, P., Ferguson, H. C., Dickinson, M. E., et al. : 1996, MNRAS 283, 1388

\bibitem[1998]{mago98}Magorrian, J., et al., 1998, AJ, 115, 2285

\bibitem[1997]{mal97}Malhotra, S., Helou, G., Stacey, G., et al.:1997, ApJ 491, L27

\bibitem[1994]{mcmahon94}McMahon R.G., Omont A., Bergeron J., Kreysa E., 
Haslam C.G.T., 1994, MNRAS, 267,  L9

\bibitem[1999]{mcmahon99}McMahon R.G., Priddey R.S., Omont A., Snellen I.,
Withington S., 1999, MNRAS, in press (astro-ph/9907239)

\bibitem[1996]{ohta96}Ohta K., Yamada T., Nakanishi K., et al.:
1996, Nature 382, 426

\bibitem[1996b]{omont96b}Omont A., McMahon R.G., Cox P., et al.: 1996b, A\&A 315, 1

\bibitem[1996a]{omont96a}Omont A., Petitjean P., Guilloteau S., McMahon R.G., Solomon P.M.:
1996a, Nature 382, 428

\bibitem[1999]{papa99}Papadopoulos P.P., Rottgering H.J.A., van der Werf P.P., et al. 1999,
ApJ, preprint (astro-ph/9908286)

\bibitem[1985]{peacock85}Peacock J.A.: 1985, MNRAS 217, 601

\bibitem[1998]{pettini98} Pettini M., Kellog M., Steidel C.C. et al.: 1998, ApJ 508, 539

\bibitem[1974]{ps74}Press W.H., Schechter P.: 1974, ApJ 187, 425

\bibitem[1996]{puget96}Puget, J.-L., Abergel, A., Bernard, J.-P., et al. : 1996, A\&A 308, L5

\bibitem[1999]{rich99}Richards E.R.: 1999, ApJ 513, L9

\bibitem[1996]{shaver96}Shaver P.A., Wall J.V., Kellermann K.I.,
Jackson C.A., Hawkins M.R.S.: 1996, Nature 384, 439

\bibitem[1993]{sco93}Scoville N.Z., Padin S., Sanders D.B. et al. : 1993, ApJ 415, L75

\bibitem[1997a]{sco97a}Scoville N.Z., Yun M.S., Bryant P.M.: 1997a, ApJ 484, 702 

\bibitem[1997b]{sco97b}Scoville N.Z., Yun M.S., Windhorst R.A., Keel W.C., Armus L.: 1997b,
ApJ 485, L21

\bibitem[1996]{shade96}Schade, D., Lilly, S. J., Le Fevre, O., Hammer, F.,
 Crampton, D.: 1996, ApJ 464, 79

\bibitem[1997]{silk97}Silk J., Spaans M.: 1997, ApJ 488, L79

\bibitem[1992]{sol92}Solomon P.M., Downes D., Radford S.J.E.: 1992, Nature 356, 318

\bibitem[1997]{sol97}Solomon P.M., Downes D., Radford S.J.E., Barrett J.W.: 1997,
  ApJ 478, 144

\bibitem[1990]{sol90}Solomon P.M., Radford S.J.E., Downes D.: 1990, ApJ 348, L53

\bibitem[1997]{smail97}Smail, I., Ivison, R.J., Blain, A.W., 1997, ApJ, 490, L5

\bibitem[1999]{smail99}Smail, I., Ivison, R.J., Kneib, J-P., et al.: 1999, MNRAS in press 
(astro-ph/9905246)

\bibitem[1996]{steidel96}Steidel, C. C., Giavalisco, M., Pettini, M., Dickinson, M.,
 Adelberger, K. L.: 1996, ApJ 462, L17

\bibitem[1995]{stein95}Stein W.A.: 1995, AJ 110, 1019

\bibitem[1999]{trentham99}Trentham N., Blain A.W., Goldader G.: 1999, MNRAS 305, 61

\bibitem[1999]{will99}Williams R.J.R., Baker A.C., Perry J.J.: 1999, MNRAS in press
(astro-ph/9907198)

\bibitem[1995]{wik95}Wiklind T., Combes F.: 1995, A\&A 299, 382 

\bibitem[1996a]{wik96a}Wiklind T., Combes F.: 1996a, Nature 379, 139  

\bibitem[1996b]{wik96b}Wiklind T., Combes F.: 1996b, A\&A 315, 86  

\bibitem[1997]{wik97}Wiklind T., Combes F.: 1997, A\&A 328, 48 

\bibitem[1997]{wink97}Wink J.E., Guilloteau S., Wilson T.L., 1997, A\&A 322, 427

\bibitem[1999]{yan99} Yan L., McCarthy P.J., Freudling W. et al.:1999, ApJ 519, L47 

\end{thebibliography}
\end{document}